\documentclass[12pt]{iopart}
\newcommand{\nn}{\nonumber}
\eqnobysec

\begin{document}

\title[Hyperbolic systems including dynamical lapse and shift]
{First-order symmetrizable
hyperbolic formulations of Einstein's equations including lapse and shift
as dynamical fields}
\author{Kashif Alvi}
\address{Theoretical Astrophysics, California Institute of Technology,
Pasadena, California 91125, USA}

\begin{abstract}
First-order hyperbolic systems are promising as a basis for numerical
integration of Einstein's equations.  In previous work, the lapse and shift
have typically not been considered part of the hyperbolic system and have been
prescribed independently.  This can be expensive computationally, especially
if the prescription involves solving elliptic equations.  Therefore,
including the lapse
and shift in the hyperbolic system could be advantageous for numerical work.
In this paper, two first-order symmetrizable hyperbolic systems are presented
that include the lapse and shift as dynamical fields and have only physical
characteristic speeds.
\end{abstract}

\pacs{04.20.-q, 04.25.Dm}

\section{Introduction}

There has been considerable interest recently in first-order hyperbolic
systems for Einstein's equations (\cite{f&r,reula,kst} and references therein).
These systems have been used in the past to prove that general relativity
has a well-posed initial value formulation \cite{f&m,friedrich1}.  Much
of the recent interest is based on the advantages that hyperbolic
formulations offer to numerical simulations \cite{friedrich2,a&y}.
The main advantage is that imposing physical boundary
conditions is much easier in the framework of a hyperbolic
system than a non-hyperbolic one.
This is especially true for boundary conditions inside a black hole
horizon \cite{friedrich2,a&y}.  Indeed, if the hyperbolic system has only
physical characteristic speeds---that is,
if the characteristic fields propagate only
on the light cones of spacetime or normal to the time slices---then the boundary condition
inside the horizon on fields propagating into the numerical grid has no effect
on the dynamics outside the horizon.\footnote{It is sufficient for the
characteristic fields to propagate on or within the light cones for this
to be true.}  Therefore, in this case, any convenient
boundary condition can be imposed inside the horizon.  This is a significant
advantage when simulating black holes.

It is particularly important to come up with stable numerical schemes
to evolve black holes since simulations of black hole collisions have an
important role to play in the detection and analysis of gravitational waves.
These simulations will be used in several stages of data analysis for
gravitational wave detectors such as the Laser Interferometer
Gravitational-Wave Observatory.  First, the simulations
are expected to yield a bank of gravitational
waveforms that will be used to detect the presence of a gravitational signal
in the detector output.  Once a signal has been detected, numerical simulations
will be used to extract binary parameters such as masses from the signal, to
test general relativity, and to do other interesting physics.

Previous numerical work has generally been restricted to systems
that do not treat the lapse and shift
as dynamical fields, but rather take them to be external to the system and
prescribe them independently.
Freedom in choosing these gauge fields
corresponds to freedom in choosing coordinates for spacetime.  This freedom can
be used for a variety of purposes, e.g., to prevent the occurrence
of coordinate singularities and reduce
coordinate shear \cite{s&y}, and to adapt the coordinate
system to the particular problem under
consideration.  When simulating black holes,
it is helpful to choose the shift
so that numerical grid points do not fall into the holes.  When
simulating binary black holes, it may
be advantageous to implement gauge conditions which
generate corotating coordinates \cite{bct,ggio}.

Some of the favored gauge choices in numerical relativity
\cite{s&y,ggio} require
solution of elliptic equations for the lapse and shift, which is expensive
computationally.
It would be more efficient to evolve the gauge fields
as part of the hyperbolic system.  However, it is important to keep some
freedom in choosing the gauge in order to allow the coordinates to be
adapted to fit specific needs.
The purpose of this paper is to present two first-order
symmetrizable hyperbolic systems which include the lapse and shift as dynamical
fields and allow four functions of spacetime to be specified freely
in the gauge prescription.

Previous work in this direction includes \cite{bmss}, in which the authors
present a weakly hyperbolic system\footnote{I refer to the full system including
lapse and shift as dynamical fields; if the shift is considered a fixed
spacetime function and not a dynamical field, then the system becomes
strongly hyperbolic.} that incorporates the gauge fields in
the system, and \cite{abpst}, in which
the authors present a new class of dynamical gauge conditions which
are not, however, part of a first-order hyberbolic system.

The first hyperbolic system presented in this paper
is based on the work of Fischer
and Marsden \cite{f&m}; it uses generalized harmonic coordinates and evolves 50
fields.  It is promising as a basis for numerical work.
The second system is based on the work of Kidder, Scheel, and
Teukolsky \cite{kst} and Lindblom and Scheel \cite{l&s}; it evolves 70 fields.
This system is not practical for numerical implementation.  Its main use is
theoretical: it allows one to show that any solution to Einstein's equations
in any gauge can be obtained using hyperbolic evolution of the entire metric,
including the gauge fields.
Both systems have only physical characteristic speeds.

In this paper, Greek indices range over 0, 1, 2, 3 and Latin indices
over 1, 2, 3.  The sign conventions are those of \cite{mtw} with $G=c=1$.
The analysis of this paper is done within the framework of a 3+1 split of
spacetime (see, e.g., \cite{mtw,yorkreview}).  In this framework, the spacetime
metric is expressed as
\begin{equation}
g_{\mu\nu}=\left(\begin{array}{cc}
-\alpha^2+\beta_k\beta^k & \beta_j \\
\beta_i & \gamma_{ij}
\end{array}\right),
\end{equation}
and the inverse 4-metric as
\begin{equation}
g^{\mu\nu}=\frac{1}{\alpha^2}\left(\begin{array}{cc}
-1 & \beta^j \\
\beta^i & \alpha^2\gamma^{ij}-\beta^i\beta^j
\end{array}\right),
\end{equation}
where $\alpha$ is the lapse, $\beta^i$ is the shift, $\gamma_{ij}$ is
the spatial 3-metric with inverse $\gamma^{ij}$, and $\beta_i=\gamma_{ij}
\beta^j$.  The unit normal to the time slices is denoted by $n^\mu$. 

I restrict attention in this paper to the vacuum Einstein equations.

\section{System I}
\subsection{Fischer-Marsden system}

Let us first briefly review the Fischer-Marsden system \cite{f&m} for
Einstein's equations.  They employ the 50 fields $g_{\mu\nu}$,
$\tilde{k}_{\mu\nu}=\partial_t g_{\mu\nu}$, and
$d_{i\mu\nu}=\partial_i g_{\mu\nu}$.  Using harmonic coordinates,
they reduce the vacuum Einstein equations $R_{\mu\nu}=0$ to the following
first-order symmetric hyperbolic system:
\begin{eqnarray}
 \partial_t g_{\mu\nu} = \tilde{k}_{\mu\nu},\nn\\
 -g^{00}\partial_t \tilde{k}_{\mu\nu} - 2g^{0i}\partial_i
	\tilde{k}_{\mu\nu} - g^{ij}\partial_i d_{j\mu\nu}
	= -2\tilde{H}_{\mu\nu},\nn\\
 g^{ij}\partial_t d_{j\mu\nu} - g^{ij}\partial_j \tilde{k}_{\mu\nu} = 0,
\label{fmfosh}
\end{eqnarray}
where $\tilde{H}_{\mu\nu}$ is a function of the fields $g_{\mu\nu},
\tilde{k}_{\mu\nu},d_{i\mu\nu}$ only and not
their derivatives.  This system is obtained by setting to zero a reduced form
of the Ricci tensor that is equal to the full Ricci tensor in
harmonic coordinates.  Using earlier work
of Choquet-Bruhat (\cite{choquet} and references therein), Fischer and Marsden
show that if the initial data for (\ref{fmfosh}) satisfy the harmonic
coordinate condition and
the constraint equations, then the solution of (\ref{fmfosh}) corresponding
to these initial data continues to
satisfy the harmonic coordinate condition
off the initial hypersurface.  Therefore, a solution of
(\ref{fmfosh}) is also a solution of the vacuum Einstein equations.

The Fischer-Marsden system (\ref{fmfosh}) has two drawbacks
when considered as a basis for
numerical integration of Einstein's equations.  The first is the restriction
to harmonic coordinates: this eliminates the freedom to choose coordinates
best suited for the physical problem at hand.  While harmonic coordinates have
been used successfully in some previous work (\cite{garfinkle} and references
therein) and are being strongly advocated for a wide variety of
applications \cite{garfinkle}, it has not yet been established whether they are
useful for simulating black hole collisions, for example.

The second drawback is that the Fischer-Marsden system has nonphysical
characteristic speeds.  As discussed above,
systems with only physical characteristic speeds are better
suited for numerical relativity, especially for black hole
simulations \cite{friedrich2,a&y}.
The characteristic speeds of the Fischer-Marsden system
can be calculated as follows: first write (\ref{fmfosh}) in the form
\begin{equation}
\partial_t u + A^i(t,x^j,u)\partial_i u = F(t,x^j,u),
\label{generalform}
\end{equation}
where $u$ is a column vector composed of the fields
($u=(g_{\mu\nu},\tilde{k}_{\mu\nu},d_{i\mu\nu})^T$
for the Fischer-Marsden system),
and the matrices $A^i$ and column
vector $F$ can depend on space and time and on the fields but not their
derivatives.  Pick a unit spatial covector $\xi_i$ (i.e., $\gamma^{ij}\xi_i
\xi_j=1$) and compute the eigenvalues
$\lambda$ of the matrix $A^i\xi_i$; $\lambda$ are the characteristic speeds
in the direction $\xi_i$.  For physical characteristic speeds, we require
$\lambda=-\beta^i\xi_i$, $-\beta^i\xi_i\pm\alpha$ (see, e.g., \cite{kst}).
However, the Fischer-Marsden system
has $\lambda=0$, $-\beta^i\xi_i\pm\alpha$.

\subsection{Generalized harmonic coordinates}

In this paper, I modify the Fischer-Marsden system to eliminate nonphysical
characteristic speeds and generalize it to include a broader range of
coordinate systems.  Let us begin by defining
$\Gamma^\mu=g^{\alpha\beta}\Gamma^\mu{}_{\alpha\beta}$
and $\Gamma_\mu=g_{\mu\nu}\Gamma^\nu$, where $\Gamma^\sigma{}_{\alpha\beta}$
are the Christoffel symbols associated with the metric $g_{\mu\nu}$ and
the coordinates $x^\mu$.
The Ricci tensor can be written as \cite{fock}
\begin{equation}
R_{\mu\nu} = \tilde{R}_{\mu\nu} + \nabla_{(\mu}\Gamma_{\nu)},
\end{equation}
where
\begin{equation}
\tilde{R}_{\mu\nu} = -\frac12 g^{\alpha\beta}\partial_\alpha\partial_\beta
	g_{\mu\nu} + H_{\mu\nu}(g,\partial g),
\end{equation}
and
\begin{equation}
H_{\mu\nu} = g^{\alpha\beta}g^{\rho\sigma}(\partial_\alpha g_{\mu\rho})
	(\partial_\beta g_{\nu\sigma})-g^{\rho\alpha}g^{\sigma\beta}
	\Gamma_{\mu\rho\sigma}\Gamma_{\nu\alpha\beta}.\label{Heqn}
\end{equation}

I generalize harmonic coordinates using Friedrich's gauge source
functions \cite{friedrich1,friedrich2} by setting
\begin{equation}
\Gamma^\mu \equiv -\nabla_\alpha\nabla^\alpha x^\mu = f^\mu(t,x^j),
\end{equation}
where the coordinates $x^\mu$ are treated as scalar fields in the
expression $\nabla_\alpha\nabla^\alpha x^\mu$, and $f^\mu$ are arbitrary
but predetermined functions of space and time.
These functions can be used to tailor the coordinates to fit specific
needs.

Consider the reduced equations obtained by setting
\begin{equation}
\tilde{R}_{\mu\nu} + \nabla_{(\mu}f_{\nu)} = 0,
\label{reducedeqn}
\end{equation}
where $f_\mu=g_{\mu\nu}f^\nu$.
Equation (\ref{reducedeqn}) will be used to write
down a first-order symmetrizable
hyperbolic system in section \ref{system1}.
Hence we must show that a solution to
(\ref{reducedeqn}) yields a solution to the vacuum Einstein equations
$R_{\mu\nu}=0$ under appropriate conditions.  I follow an argument due to
Friedrich \cite{friedrich1,friedrich2} which is based on earlier work by
Choquet-Bruhat (\cite{choquet} and references therein).

Let $g_{\mu\nu}$ be a solution to (\ref{reducedeqn}).  Compute
$\Gamma^\mu$ and $R_{\mu\nu}$ from $g_{\mu\nu}$, and let $h^\mu
=\Gamma^\mu-f^\mu$.  Then $R_{\mu\nu}=\tilde{R}_{\mu\nu} +
\nabla_{(\mu}\Gamma_{\nu)}=\nabla_{(\mu}h_{\nu)}$ where $h_\mu=\Gamma_\mu
-f_\mu$.  The Einstein tensor is
\begin{equation}
G_{\mu\nu}=R_{\mu\nu}-\frac R2 g_{\mu\nu}=\nabla_{(\mu}h_{\nu)}-\frac12
	g_{\mu\nu}\nabla_\alpha h^\alpha,
\label{Geqn}
\end{equation}
and the contracted Bianchi identities $\nabla^\mu G_\mu{}^\nu=0$ imply
\begin{equation}
\nabla^\mu\nabla_{\mu}h^\nu+R^\nu{}_\mu h^\mu=0,
\end{equation}
which is the subsidiary equation derived by Friedrich
\cite{friedrich1,friedrich2}.  Since this is a linear homogeneous wave
equation for $h^\mu$, we conclude that if $h^\mu=0$ and $\nabla_\nu h^\mu=0$
on the initial hypersurface, then $h^\mu=0$ in a neighborhood of the initial
hypersurface.  This implies $R_{\mu\nu}=\nabla_{(\mu}h_{\nu)}=0$ in this
neighborhood.  So $g_{\mu\nu}$ is a solution to the vacuum Einstein
equations in a neighborhood of the initial hypersurface.  This solution is
obtained in coordinates satisfying $\Gamma^\mu=f^\mu$.

We therefore need to ensure
\begin{eqnarray}
[\Gamma^\mu-f^\mu]_{t=0}=0,\label{initialreq1}\\
{[}\nabla_{\nu}(\Gamma^{\mu}-f^{\mu})]_{t=0}=0,\label{initialreq2}
\end{eqnarray}
where the time slice $t=0$ represents the initial hypersurface.
Given a spatial 3-metric $\gamma_{ij}$ and an
extrinsic curvature $K_{ij}$ that satisfy the constraint equations, we will
construct initial data for our system such that (\ref{initialreq1}) is
satisfied.  Equation (\ref{initialreq2}) will then follow from the constraint
equations.  This will be discussed in detail in section \ref{initdata}.

\subsection{System I}
\label{system1}

Define the fields
\begin{eqnarray}
 k_{\mu\nu}=\partial_t g_{\mu\nu}-\beta^j\partial_j g_{\mu\nu},\label{kdef}\\
 d_{i\mu\nu}=\partial_i g_{\mu\nu}.\label{ddef}
\end{eqnarray}
Here and throughout this section, $\beta^i$ will be considered convenient
shorthand for $-g^{0i}/g^{00}$, and similarly $\alpha$ for $(-g^{00})^{-1/2}$.
The new field $k_{\mu\nu}$ is a replacement for $\tilde{k}_{\mu\nu}$ and has
been introduced to eliminate nonphysical characteristic speeds.

The first-order symmetrizable hyperbolic system presented in this section
is based on the 50 fields $g_{\mu\nu}$,
$k_{\mu\nu}$, and $d_{i\mu\nu}$.  The definition (\ref{kdef}) yields
an expression for $\partial_t g_{\mu\nu}$ in terms of the 50 fields and their
first spatial derivatives.  An expression for $\partial_t d_{i\mu\nu}$
is obtained through equality of mixed partials: $\partial_t d_{i\mu\nu}=
\partial_i\partial_t g_{\mu\nu}=\partial_i(k_{\mu\nu}+\beta^j d_{j\mu\nu})$.
Finally, an expression for $\partial_t k_{\mu\nu}$ is obtained from the
reduced equation (\ref{reducedeqn}).  To summarize, we have the first-order
system
\begin{eqnarray}
\fl \partial_t g_{\mu\nu}+\frac{g^{0i}}{g^{00}}\partial_i g_{\mu\nu}
	=k_{\mu\nu},\label{geqn}\\
\fl \partial_t k_{\mu\nu}+\frac{g^{0i}}{g^{00}}\partial_i k_{\mu\nu}+\frac
	{\gamma^{ij}}{g^{00}}\partial_i d_{j\mu\nu}=-\frac{\gamma^{ij}}{g^{00}}
	g^{0\alpha}d_{i\mu\nu}k_{\alpha j}+\frac{2}{g^{00}}[H_{\mu\nu}
+\partial_{(\mu}f_{\nu)}-\Gamma^\alpha{}_{\mu\nu}f_\alpha],\label{keqn}\\
\fl \partial_t d_{i\mu\nu}+\frac{g^{0j}}{g^{00}}\partial_j d_{i\mu\nu}
	-\partial_i k_{\mu\nu}=\frac{\gamma^{jk}}{g^{00}}g^{0\alpha}d_{j\mu\nu}
	d_{i\alpha k},\label{deqn}
\end{eqnarray}
where $\gamma^{ij}=(g^{00})^{-2}(g^{00}g^{ij}-g^{0i}g^{0j})$ is the inverse of
the 3-metric $\gamma_{ij}$.  In (\ref{keqn}), $H_{\mu\nu}$ is to be expressed
via (\ref{Heqn}) in terms of the fields only and not their derivatives [using
(\ref{kdef}) and (\ref{ddef})].  In addition, in
(\ref{geqn})--(\ref{deqn}), the inverse 4-metric is considered to be a function
of $g_{\mu\nu}$ and not a fundamental field.  In deriving these expressions,
I have used the relation
\begin{equation}
\partial_\alpha g^{\mu\nu}=-g^{\mu\theta}g^{\nu\lambda}\partial_\alpha
	g_{\theta\lambda}.
\end{equation}
The system (\ref{geqn})--(\ref{deqn}) will be called system I.

\subsection{Initial data}
\label{initdata}

It remains to specify how to set initial data for system I to ensure
(\ref{initialreq1}) and (\ref{initialreq2}) are satisfied.  Begin with a
solution $(\gamma_{ij},K_{ij})$ of the constraint equations, where $K_{ij}$
represents the extrinsic curvature of the initial hypersurface.  First set
$g_{ij}=\gamma_{ij}$.  We are free to choose $g_{0\mu}$ on the initial
hypersurface as long as $g_{00}<g_{0i}g_{0j}\gamma^{ij}$.  This requirement
is equivalent to $\alpha^2>0$ and implies $g^{00}<0$.  Freedom in choosing
$g_{0\mu}$ corresponds to freedom in choosing the lapse and shift at $t=0$.

We now have $g_{\mu\nu}|_{t=0}$.  Next set $d_{i\mu\nu}=\partial_i g_{\mu\nu}
|_{t=0}$.  The final step is to fill in $k_{\mu\nu}$ from $K_{ij}$ and the
requirement (\ref{initialreq1}).  The extrinsic curvature can be expressed as
\begin{equation}
K_{ij}=-\frac{1}{2\alpha}(\partial_t\gamma_{ij}-\beta^k\partial_k\gamma_{ij}
	-2\gamma_{k(i}\partial_{j)}\beta^k).
\label{Kdef}
\end{equation}
From this we deduce
\begin{equation}
k_{ij}=-2\alpha K_{ij}+2g_{k(i}\partial_{j)}\beta^k,
\end{equation}
which can be used to fill in $k_{ij}$ at $t=0$.

The quantities $k_{0\mu}$ are obtained from the requirement
(\ref{initialreq1}).
Writing out $\Gamma^\mu$ in terms of the metric and its first derivatives,
we obtain
\begin{eqnarray}
\fl \Gamma^0 = -\alpha^{-3}(\partial_t\alpha-\beta^i\partial_i\alpha
	+\alpha^2 K),\\
\fl \Gamma^i = -\alpha^{-2}(\partial_t\beta^i-\beta^j\partial_j\beta^i)
	+\alpha^{-3}(\partial_t\alpha-\beta^j\partial_j\alpha
+\alpha^2 K)\beta^i-\alpha^{-1}\gamma^{ij}\partial_j\alpha+{}^{(3)}
	\Gamma^i{}_{jk}\gamma^{jk},
\end{eqnarray}
where $K=\gamma^{ij}K_{ij}$, and ${}^{(3)}\Gamma^i{}_{jk}$ are the Christoffel
symbols associated with the 3-metric $\gamma_{ij}$ and the spatial coordinates
$x^j$.  Setting $\Gamma^\mu=f^\mu$ gives us expressions for $\partial_t\alpha$
and $\partial_t\beta^i$ which we use to fill in $k_{0\mu}$ at $t=0$:
\begin{eqnarray}
 k_{0i}=B_i+\beta^j k_{ij},\\
 k_{00}=2\alpha^3(\alpha f^0+K)+2\beta^i B_i+\beta^i\beta^j k_{ij},
\end{eqnarray}
where
\begin{equation}
B_i=-\alpha^2(g_{i\mu}f^\mu+\alpha^{-1}\partial_i\alpha-{}^{(3)}\Gamma_{ijk}
	\gamma^{jk}).
\end{equation}

The initial data for system I is now complete and
satisfies the constraint equations
\begin{equation}
G_{\mu\nu}n^\nu|_{t=0}=0
\label{constraints}
\end{equation}
and the requirement (\ref{initialreq1}).
This in fact implies that the requirement
(\ref{initialreq2}) is satisfied.  The argument follows earlier work
\cite{choquet} on the reduction of Einstein's equations using harmonic
coordinates.  From (\ref{Geqn}) and (\ref{constraints}), we deduce
\begin{equation}
2n^\nu\nabla_{(\mu}h_{\nu)}-n_\mu\nabla_\alpha h^\alpha=0.
\label{gradh}
\end{equation}
Here and in the remainder of the paragraph, all quantities are evaluated at
$t=0$.  We know $h^\mu\equiv\Gamma^\mu-f^\mu=0$ on the initial hypersurface,
so $v^\nu\nabla_\nu h^\mu=0$
for any spatial vector $v^\mu$ (i.e., for $v^\mu$ satisfying $v^\mu n_\mu=0$).
It remains to show $n^\nu\nabla_\nu h^\mu=0$.  By contracting (\ref{gradh})
with $v^\mu$, we obtain $v^\mu n^\nu\nabla_\nu h_\mu=0$.  Furthermore,
$\nabla_\alpha h^\alpha=-n^\mu n^\nu\nabla_\mu h_\nu$.  Contracting
(\ref{gradh}) with $n^\mu$, we obtain $n^\mu n^\nu\nabla_\mu h_\nu=0$.
It follows that
$n^\nu\nabla_\nu h^\mu=0$ and so (\ref{initialreq2}) is satisfied.

Therefore, a solution $(g_{\mu\nu},k_{\mu\nu},d_{i\mu\nu})$ to system I
with initial data as constructed above yields a
solution $g_{\mu\nu}$ to the vacuum Einstein equations.

\subsection{Hyperbolicity of system I}

System I is symmetrizable hyperbolic.  To see this,
let $u=(g_{\mu\nu},k_{\mu\nu},d_{1\mu\nu},d_{2\mu\nu},d_{3\mu\nu})^T$
and write equations
(\ref{geqn})--(\ref{deqn}) in the form (\ref{generalform}).  This determines
the $50\times 50$ matrices $A^i$ to be
\begin{equation}
A^i=\left(\begin{array}{ccccc}
-\beta^i I & 0 & 0 & 0 & 0 \\
0 & -\beta^i I & -\alpha^2\gamma^{i1}I & -\alpha^2\gamma^{i2}I &
-\alpha^2\gamma^{i3}I \\
0 & -\delta_1{}^i I & -\beta^i I & 0 & 0 \\
0 & -\delta_2{}^i I & 0 & -\beta^i I & 0 \\
0 & -\delta_3{}^i I & 0 & 0 & -\beta^i I \end{array}\right).
\end{equation}
Here and in equations (\ref{symmetrizer})--(\ref{V}),
0 is the $10\times 10$ zero matrix
and $I$ is the $10\times 10$ identity matrix.  It can be checked easily that
the positive definite symmetric $50\times 50$ matrix
\begin{equation}
H=\left(\begin{array}{ccccc}
I & 0 & 0 & 0 & 0 \\
0 & \alpha^{-2}I & 0 & 0 & 0 \\
0 & 0 & \gamma^{11}I & \gamma^{12}I & \gamma^{13}I \\
0 & 0 & \gamma^{12}I & \gamma^{22}I & \gamma^{23}I \\
0 & 0 & \gamma^{13}I & \gamma^{23}I & \gamma^{33}I \end{array}\right),
\label{symmetrizer}
\end{equation}
is a symmetrizer for the system, i.e., $HA^i$ are symmetric matrices.

Moreover, system I has only physical characteristic speeds; that is,
the eigenvalues of
$A^i\xi_i$ are $\lambda_+ = -\beta^i\xi_i+\alpha$, $\lambda_0 = -\beta^i\xi_i$,
and $\lambda_- = -\beta^i\xi_i-\alpha$.  Let $\eta_i$ and $\chi_i$ be
unit spatial covectors that form an orthonormal triad with $\xi_i$,
that is, $\eta^i\xi_i=0=\chi^i\xi_i=\eta^i\chi_i$ and
$\eta^i\eta_i=1=\chi^i\chi_i$, where $\eta^i=\gamma^{ij}\eta_j$ and
$\chi^i=\gamma^{ij}\chi_j$.
Let us construct a $50\times 50$ matrix $V$ whose columns are 50 linearly
independent eigenvectors of $A^i\xi_i$.  One such matrix is
\begin{equation}
V=\left(\begin{array}{ccccc}
0 & I & 0 & 0 & 0 \\
-\alpha I & 0 & 0 & 0 & \alpha I \\
\xi_1 I & 0 & \eta_1 I & \chi_1 I & \xi_1 I \\
\xi_2 I & 0 & \eta_2 I & \chi_2 I & \xi_2 I \\
\xi_3 I & 0 & \eta_3 I & \chi_3 I & \xi_3 I \end{array}\right).
\label{V}
\end{equation}
The first ten columns of $V$ are eigenvectors of $A^i\xi_i$
with eigenvalue $\lambda_+$; the next thirty columns have eigenvalue
$\lambda_0$; and the last ten columns have eigenvalue $\lambda_-$.
The characteristic fields in the direction $\xi_i$ are obtained from
$V^{-1}u$ and are given by $\alpha^{-1}k_{\mu\nu}\pm \xi^i d_{i\mu\nu}$,
$g_{\mu\nu}$, $\eta^i d_{i\mu\nu}$, and $\chi^i d_{i\mu\nu}$.

\section{System II}

In this section, all indices are lowered and raised by the spatial 3-metric
$\gamma_{ij}$ and its inverse $\gamma^{ij}$.
The second system presented in this paper is based on a hyperbolic system
in \cite{kst}, which is in turn based on the ADM equations \cite{adm}.  The
system in \cite{kst}, called system 1, employs the 30 fields
$\gamma_{ij}$, $K_{ij}$, and
\begin{equation}
d_{kij}=\partial_k\gamma_{ij}.
\label{dkijdef}
\end{equation}
It is obtained by densitizing the lapse and adding multiples of the constraint
equations to the evolution equations.  The relevant constraints are the
Hamiltonian constraint
\begin{equation}
C=\frac12({}^{(3)}R-K_{ij}K^{ij}+K^2)=0,
\end{equation}
the momentum constraints
\begin{equation}
C_i=D_j K_i{}^j-D_i K=0,
\end{equation}
and the constraint
\begin{equation}
C_{ijkl}=\partial_{[i}d_{j]kl}=0,
\end{equation}
where ${}^{(3)}R$ and $D_i$ are the Ricci scalar and covariant derivative
associated with $\gamma_{ij}$, and $K=\gamma^{ij}K_{ij}$.

System 1 has five free parameters
that govern how to densitize the lapse and how much of the constraints to add;
these parameters determine the system's hyperbolicity.  In
fact, it has been shown \cite{l&s} that for a certain range of these
parameters, system 1 is symmetrizable hyperbolic and has only physical
characteristic speeds.

Here I construct a first-order symmetrizable hyperbolic system based on system
1 that includes
the lapse and shift in the system.  Let us begin by defining the
densitized lapse
\begin{equation}
Q=\ln(\alpha\gamma^{-1/2}),
\label{Qdef}
\end{equation}
where $\gamma=\det(\gamma_{ij})$.  Next define the new fields
\begin{eqnarray}
Q_i=\partial_i Q, \qquad Q_{ij}=\partial_i\partial_j Q,\nn\\
b_i{}^j=\partial_i\beta^j, \qquad b_{ij}{}^k=\partial_i\partial_j\beta^k.
\label{newfields}
\end{eqnarray}
Note that $Q_{ij}=Q_{(ij)}$ and $b_{ij}{}^k=b_{(ij)}{}^k$.  The hyperbolic
system presented in this section is based on the 70 fields $\gamma_{ij},
K_{ij},d_{kij},Q,Q_i,Q_{ij},\beta^i,b_i{}^j,b_{ij}{}^k$.

Expressions for time derivatives of these fields are obtained as follows.
First, $\partial_t\gamma_{ij}$ is obtained from (\ref{Kdef}):
\begin{equation}
\partial_t\gamma_{ij}-\beta^k\partial_k\gamma_{ij}=-2\alpha K_{ij}
	+2\gamma_{k(i}b_{j)}{}^k.
\label{gammaeqn}
\end{equation}
This is one of the ADM evolution equations with the new
fields (\ref{newfields}) substituted in.
Here and henceforth, it is understood that $\alpha$
is to be rewritten in terms of $Q$ using (\ref{Qdef}).  Following \cite{kst},
I add
$\zeta_1\alpha\gamma_{ij}C$ and $\zeta_2\alpha\gamma^{mn}C_{m(ij)n}$ to
the second ADM evolution equation (which is equation (2.9) in \cite{kst}),
where $\zeta_1$ and $\zeta_2$ are free
parameters.  Rewriting this equation in terms of the new fields
(\ref{newfields}), we obtain
\begin{eqnarray}
\fl \partial_t K_{ij}=\beta^k\partial_k K_{ij}-\frac12\alpha\gamma^{mn} [
	\partial_m d_{nij}+2\partial_{(i}d_{j)mn}-(1-\zeta_2)\partial_{(i}
	d_{|mn|j)}-(1+\zeta_2)\partial_m d_{(ij)n}\nn\\
{}-\zeta_1\gamma_{ij}
	\gamma^{kl}(\partial_m d_{kln}-\partial_k d_{lmn})]
	+ 2K_{k(i}b_{j)}
	{}^k-\alpha[2K_{im}K^m{}_j-KK_{ij}+Q_{ij}\nn\\
{}+(d_{(ij)m}-\frac12 d_{mij})
	(\tilde{d}^m-d^m-Q^m)+d^{mn}{}_i d_{[nm]j}-\frac34 d_{imn}d_j{}^{mn}
	+Q_i Q_j\nn\\
{}+Q_{(i}d_{j)}+\frac14 d_i d_j]+\frac12 \zeta_1\alpha\gamma_{ij}
	(\tilde{d}_m d^m-\tilde{d}_m\tilde{d}^m-\frac14 d_m d^m-\frac12 d_{klm}
	d^{mkl}\nn\\
{}+\frac34 d_{klm}d^{klm}-K_{mn}K^{mn}+K^2),
\label{Keqn}
\end{eqnarray}
where $d_i=\gamma^{jk}d_{ijk}$ and $\tilde{d}_i=\gamma^{jk}d_{jki}$. 

Using equality of mixed partials, we have $\partial_t d_{kij}=\partial_k
\partial_t\gamma_{ij}$ which, together with a spatial derivative
of (\ref{gammaeqn}), yields an
evolution equation for $d_{kij}$.  Following \cite{kst}, I add
$\zeta_3\alpha\gamma_{k(i}C_{j)}$ and $\zeta_4\alpha\gamma_{ij}C_k$ to this
equation and use (\ref{newfields}) to obtain
\begin{eqnarray}
\fl \partial_t d_{kij} = \beta^m\partial_m d_{kij}
	+\alpha\gamma^{mn} [\zeta_3(\gamma_{k(i}\partial_{|m}K_{n|j)}
	-\gamma_{k(i}\partial_{j)}K_{mn})
	+\zeta_4\gamma_{ij}(\partial_m K_{nk}-\partial_k K_{mn})]\nn\\
{}-2\alpha\partial_k K_{ij}+2\gamma_{m(i}b_{j)k}{}^m+d_{mij}b_k{}^m
	+2d_{km(i}b_{j)}{}^m\nn\\
{}-\alpha K_{ij}(2Q_k+d_k)+\alpha\zeta_4\gamma_{ij}
	[K_{km}(\frac12 d^m-\tilde{d}^m)+\frac12 K^{mn}d_{kmn}]\nn\\
{}+\alpha\zeta_3[\gamma_{k(i}K_{j)m}(\frac12 d^m-\tilde{d}^m)
	+\frac12 K^{mn}\gamma_{k(i}d_{j)mn}],
\label{dkijeqn}
\end{eqnarray}
where $\zeta_3$ and $\zeta_4$ are free parameters.
The parameters $(\zeta_1,\zeta_2,\zeta_3,\zeta_4)$ in the above equations
correspond to the parameters $(\gamma,\zeta,\eta,\chi)$ in \cite{kst}.
The parameter $\sigma$ in \cite{kst} has been set to 1/2 by
the definition (\ref{Qdef}).

The next step is to specify evolution equations for the lapse density and
shift.  Spatial derivatives of these equations will
then yield evolution equations for the fields
(\ref{newfields}).  I consider a particular form for the lapse density and
shift evolution equations, a form that results in a symmetrizable hyperbolic
system but yet allows four functions of spacetime to be freely specified.
The equations are
\begin{eqnarray}
\partial_t Q-\beta^i\partial_i Q = \psi^0(t,x^j;Q),\label{Qeqn}\\
\partial_t\beta^i-\beta^j\partial_j\beta^i = \psi^i(t,x^k;Q,\beta^m),
\label{betaeqn}
\end{eqnarray}
where $\psi^\mu$
are arbitrary but predetermined functions of space,
time, and lapse density (and of shift in the case of $\psi^i$).

Evolution equations for the fields (\ref{newfields}) are obtained by taking
spatial derivatives of (\ref{Qeqn}) and (\ref{betaeqn}), and using equality
of mixed partials.  For example, $\partial_t Q_i=\partial_i\partial_t Q=
\partial_i(\beta^j\partial_j Q+\psi^0)$.  We obtain
\begin{eqnarray}
\partial_t Q_i-\beta^j\partial_j Q_i = Q_j b_i{}^j+\partial_i\psi^0,\label{Qieqn}\\
\partial_t Q_{ij}-\beta^k\partial_k Q_{ij} = 2Q_{k(i}b_{j)}{}^k+Q_k b_{ij}{}^k
	+\partial_i\partial_j \psi^0,\label{Qijeqn}\\
\partial_t\beta_i{}^j-\beta^k\partial_k\beta_i{}^j = b_i{}^k b_k{}^j+\partial_i
	\psi^j,\label{bijeqn}\\
\partial_t b_{ij}{}^k-\beta^m\partial_m b_{ij}{}^k = 2b_{(i}{}^m b_{j)m}{}^k
	+b_{ij}{}^m b_m{}^k+\partial_i\partial_j \psi^k,\label{bijkeqn}
\end{eqnarray}
where it is understood that the spatial
derivatives of $\psi^\mu$ are to be
written, using (\ref{newfields}), in terms of fields only and not derivatives
of fields.

When the system (\ref{gammaeqn})-(\ref{bijkeqn}), called system II,
is put in the form
(\ref{generalform}) with $u=(\gamma_{ij},K_{ij},d_{kij},Q,Q_i,Q_{ij},\beta^i,
b_i{}^j,b_{ij}{}^k)^T$, the $70\times 70$ matrices $A^i$ have the block
diagonal form
\begin{equation}
A^i=\left(\begin{array}{cc}
\tilde{A}^i_{30\times 30} & 0_{30\times 40} \\
0_{40\times 30} & -\beta^i I_{40\times 40} \end{array}\right).
\label{Ai}
\end{equation}
The nontrivial parts $\tilde{A}^i$ of $A^i$ come from the evolution equations
(\ref{gammaeqn})-(\ref{dkijeqn}) for the 30 fields
$\gamma_{ij},K_{ij},d_{kij}$.
Since the principal parts of these equations
are identical (after relabeling the free parameters as indicated above)
to the principal parts of the system 1 evolution equations for
$\gamma_{ij},K_{ij},d_{kij}$ given in \cite{kst},
the matrices $\tilde{A}^i$ are identical to the corresponding
matrices in \cite{kst}.
This implies that if system 1 is symmetrizable,
so is system II.  Indeed, the matrix
\begin{equation}
H=\left(\begin{array}{cc}
\tilde{H}_{30\times 30} & 0_{30\times 40} \\
0_{40\times 30} & I_{40\times 40} \end{array}\right),
\end{equation}
where $\tilde{H}_{30\times 30}$ symmetrizes system 1, is a symmetrizer for
system II.  In other words, if the $30\times 30$ matrices
$\tilde{H}\tilde{A}^i$ are symmetric, then so are the $70\times 70$
matrices $HA^i$.
In addition, the characteristic fields $Q,Q_i,Q_{ij},
\beta^i,b_i{}^j,b_{ij}{}^k$ all propagate normal
to the time slices.

It has been shown \cite{l&s} that system 1 in \cite{kst} is symmetrizable
and has only physical characteristic speeds when the free parameters are chosen
as follows:
\begin{eqnarray}
\zeta_3=\frac{-8}{5+10\zeta_1+7\zeta_2+6\zeta_1\zeta_2},\qquad &
\zeta_4=-\frac{4+10\zeta_1+4\zeta_2+6\zeta_1\zeta_2}
	{5+10\zeta_1+7\zeta_2+6\zeta_1\zeta_2},\nn\\
-5/3 < \zeta_2 < 0, &
5+10\zeta_1+7\zeta_2+6\zeta_1\zeta_2 \neq 0.
\label{zetaeqns}
\end{eqnarray}
We conclude that for the same choice of parameters, system II is symmetrizable
and has only physical characteristic speeds.

System II is not practical for numerical implementation.  Since the lapse
density and shift evolution equations (\ref{Qeqn}) and (\ref{betaeqn}) decouple
from the rest of the system, they can be evolved separately to obtain the
lapse density and shift as spacetime functions.  These functions can then be
substituted into system 1 in \cite{kst}.  Therefore, the full
seventy-field system II does not need to be evolved; the thirty-field system 1
suffices.

However, system II is useful from a theoretical point of view.  Consider a
solution of Einstein's equations in an arbitrary
gauge.  Using the densitized lapse and shift from this solution,
compute the left-hand sides of equations (\ref{Qeqn}) and (\ref{betaeqn}).
Set the
spacetime functions $\psi^\mu$ equal to these computed quantities.
Take initial values for the fields in system II from the spacetime metric
under consideration.
System II can now be used, with these initial values and with $\psi^\mu$
as defined above, to obtain the entire metric
by evolving hyperbolic equations that are part of a symmetrizable system
with only
physical characteristic speeds.  So system II can be used to obtain any
solution of Einstein's equations in any gauge using hyperbolic evolution
for the entire metric, including the densitized lapse and shift.  Note,
however, that the lapse is not evolved directly in this system; it is
obtained from the densitized lapse via equation (\ref{Qdef}).

\section{Future directions}

An important future research direction
is to study and understand the stability of numerical implementations of
system I.  It has been shown in previous
work \cite{kst}
that some hyperbolic systems are more stable than others when used
to simulate black holes in three spatial dimensions.  The reasons for
this behavior are not yet understood.
Another future research direction is to explore how to use the
free functions $f^\mu$ in system I to control the coordinate system.

\ack

I am grateful to Lee Lindblom and Mark Scheel for valuable discussions
and for sharing their results before publication, and to Olivier Sarbach,
Kip Thorne, and Manuel Tiglio for useful comments on the manuscript.
This research was supported in part by NSF grant PHY-9900776.

\end{document}